\documentclass[showpacs,showkeys,aps,prb,floatfix,reprint,superscriptaddress,footinbib,citeautoscript]{revtex4-2}
\expandafter\def\csname figurename\endcsname{Figure}

\usepackage{amsmath}
\usepackage{booktabs}
\usepackage{xcolor}
\usepackage{soul}
\usepackage{graphicx}
\usepackage{hyperref}
\usepackage{microtype}
\usepackage{multirow}
\usepackage{placeins}
\usepackage{setspace}
\usepackage{subcaption}
\usepackage{ragged2e}
\usepackage{siunitx}
\usepackage{orcidlink}
\usepackage{chemmacros}

\usepackage{chemmacros}
\usepackage{color}
\usepackage{hyperref} \hypersetup{colorlinks=true,citecolor=blue,linkcolor=blue,urlcolor=blue}

\makeatletter
\renewcommand\frontmatter@abstractwidth{\dimexpr\textwidth\relax}
\makeatother

\def\NCSU{\footnotesize Department of Materials Science and Engineering, North Carolina State University, Raleigh, NC 27695-7907, USA}
\def\MST{\footnotesize Department of Materials Science and Engineering, Missouri University of Science and Technology, Rolla, MO 65409, USA}
\def\DUKEMEMS{Department of Mechanical Engineering and Materials Science, Duke University, Durham, NC 27708, USA}
\def\DUKECEM{Center for Extreme Materials, Duke University, Durham, NC 27708, USA}

\begin{document}

\title{Grain-Boundary Premelting in High-Entropy Transition Metal Carbides}

\author{Marium~M.~Mou\,\orcidlink{0009-0008-4839-5714}}
\affiliation{\NCSU}
\author{Caleb~Schenck\,\orcidlink{0009-0009-5191-8042}}
\affiliation{\NCSU}
\author{Samuel~E.~Daigle\,\orcidlink{0000-0002-8892-4237}}
\affiliation{\NCSU}
\author{William~G.~Fahrenholtz\,\orcidlink{0000-0002-8497-0092}}
\affiliation{\MST}
\author{Bharat~Gwalani\,\orcidlink{0000-0002-3021-6676}}
\affiliation{\NCSU}
\author{Stefano~Curtarolo\,\orcidlink{0000-0003-0570-8238}}
\affiliation{\DUKEMEMS}\affiliation{\DUKECEM}
\author{Donald~W.~Brenner\,\orcidlink{0009-0009-1618-4469}}
\email[]{brenner@ncsu.edu}
\affiliation{\NCSU}

\date[]{}

\begin{abstract}
\noindent
Grain-boundary segregation and thermally induced interfacial disordering were investigated in four high-entropy transition metal carbides using Monte Carlo (MC) sampling and molecular dynamics (MD) with the universal MACE-OMAT-0 machine-learning interatomic potential. MC sampling segregated the group-VI element (Cr, Mo, or W) and Zr to grain boundaries, where the group-VI content reached approximately 45~at.\%, consistent with STEM--EDS observations. During MD heating, the grain-boundary Lindemann index, a normalized measure of interatomic distance fluctuations, reached the liquid-like threshold of $\delta=0.15$ near $1390^{\circ}\mathrm{C}$ for the Cr-containing carbides, $1660^{\circ}\mathrm{C}$ for Mo, and $1890^{\circ}\mathrm{C}$ for W, while the grain interiors remained below the threshold. A chemically random \ch{(Cr,Hf,Ta,Ti,Zr)C} reference crossed about $60^{\circ}\mathrm{C}$ later and showed less boundary-localized disorder, highlighting the role of interfacial chemistry in premelting. Species-resolved displacements showed enhanced grain-boundary mobility, particularly for carbon. Overall, Cr-rich interfaces showed the earliest and most extensive premelting-like response, followed by Mo- and W-containing boundaries.
\end{abstract}


\maketitle

\section{Introduction}
Early studies of high-entropy transition metal carbides (HETMCs) emphasized phase stability and bulk property predictions using descriptors such as entropy-forming ability and valence electron concentration~\cite{sarker_high-entropy_2018,hossain_entropy_2021,divilov_disordered_2024_etal}. However, as studies of these materials have matured, attention has increasingly shifted toward processing-dependent microstructures, grain boundaries (GBs), and their effects on mechanical behavior.

Among the group-VI transition metals \ch{(Cr,Mo,W)}, chromium occupies a unique position. Several recent experimental studies have reported Cr segregation to GBs in HETMCs and the effect of this segregation on mechanical properties. Wang et al.~\cite{wang_role_2022}, for example, characterized \ch{(Cr,Hf,Nb,Ta,Zr)C} prepared using spark plasma sintering as a function of the initial relative Cr precursor. This composition is predicted to be multi-phase based on the Disordered Enthalpy–Entropy Descriptor (DEED)~\cite{divilov_disordered_2024_etal}. They report limited Cr solubility of about 2.5~at.\% in the grains, Cr enrichment at the GBs, and a higher density, which they suggest is related to relatively low melting temperature of Cr carbide phases. In another series of studies, Su et al.~\cite{su_insights_2023} characterized the solubility and segregation of Cr in \ch{(Cr,Nb,Ta,Ti,Zr)C} created by pressureless sintering at 1800--2000$^\circ$C. The DEED value of 20 for this composition predicts a single-phase, but the value is near the single-multiphase threshold of 19. They report a Cr solubility limit of $\sim$3.8 at.\%, Cr segregation to GBs, and formation of $\mathrm{Cr}_3\mathrm{C}_2$ at triple junctions as the Cr concentration is increased.  In a subsequent study, this composition was processed via hot-pressing at 2100$^\circ$C using powder synthesized from carbothermal reduction~\cite{su_fracture_2024}. They report no second phases and a decrease in porosity with Cr addition. They attribute this to flow of liquid Cr carbide inside and out of the material during processing. They also report a transition from inter- to trans-granular fracture with increasing Cr content, suggesting an increase in the GB strength. They suggest that this may be due to a combination of impurities that are expelled with the flow of liquid Cr carbide out of the sample during hot pressing and a lattice mismatch at the grain boundaries that depends on Cr concentration. Whether this behavior is unique to Cr remains unclear. Group-VI transition
metals exhibit markedly different carbide thermodynamics and melting
temperatures, suggesting that the stability of a chemically segregated
GB may depend strongly on the identity of the segregating
species.

{Schenck~et~al.~\cite{schenck_effect_2026}} studied the structure and mechanical response of \ch{(Cr,Hf,Ta,Ti,Zr)C} synthesized using carbothermal reduction of metal oxides with carbon. The resulting powders were densified in mild vacuum using direct current sintering at  2000$^\circ$C and a peak uniaxial pressure of 50 MPa. This composition has a DEED value of 18 and is therefore predicted to be multi-phase~\cite{divilov_disordered_2024_etal}. Like prior work, they report $<5~\mathrm{at.}\%$ Cr concentration in the grains and strong Cr segregation to the GBs. Schenck et al.\ also studied the composition
\ch{(Hf,Ta,Ti,W,Zr)C}, in which W replaces Cr. Similar to the
Cr-containing composition, they observed a reduced concentration of W in the bulk
and segregation of W to the grain boundaries. However, the W-containing composition
retained a larger bulk concentration of W, approximately 11.2~at.\%, compared with less
than 5~at.\% Cr in the Cr-containing composition, and exhibited a noticeably lower
interfacial concentration of W than the corresponding interfacial concentration of Cr.

\begin{figure*}[htb!]
    {
    \begin{subfigure}[b]{0.32\textwidth}
        \includegraphics[width=\linewidth]{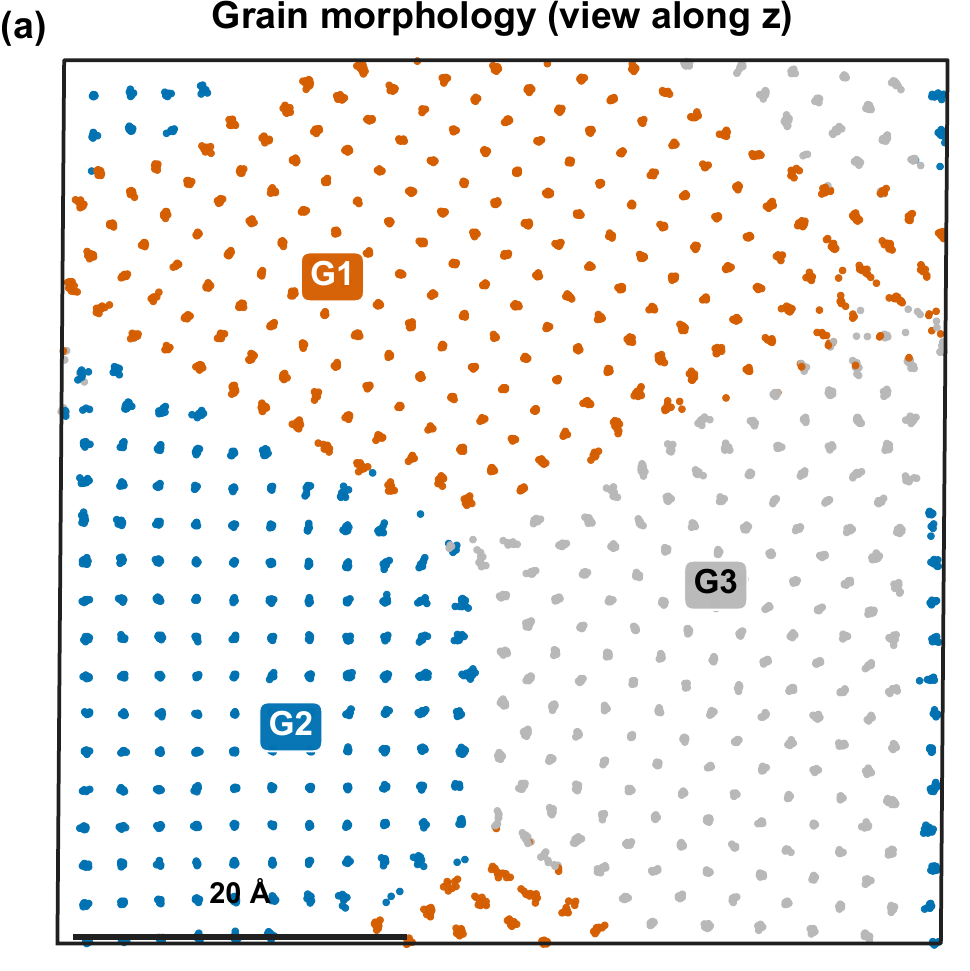}
        \phantomsubcaption
        \label{fig:hec_morphology}
    \end{subfigure}
    \hfill
    \begin{subfigure}[b]{0.32\textwidth}
        \includegraphics[width=\linewidth]{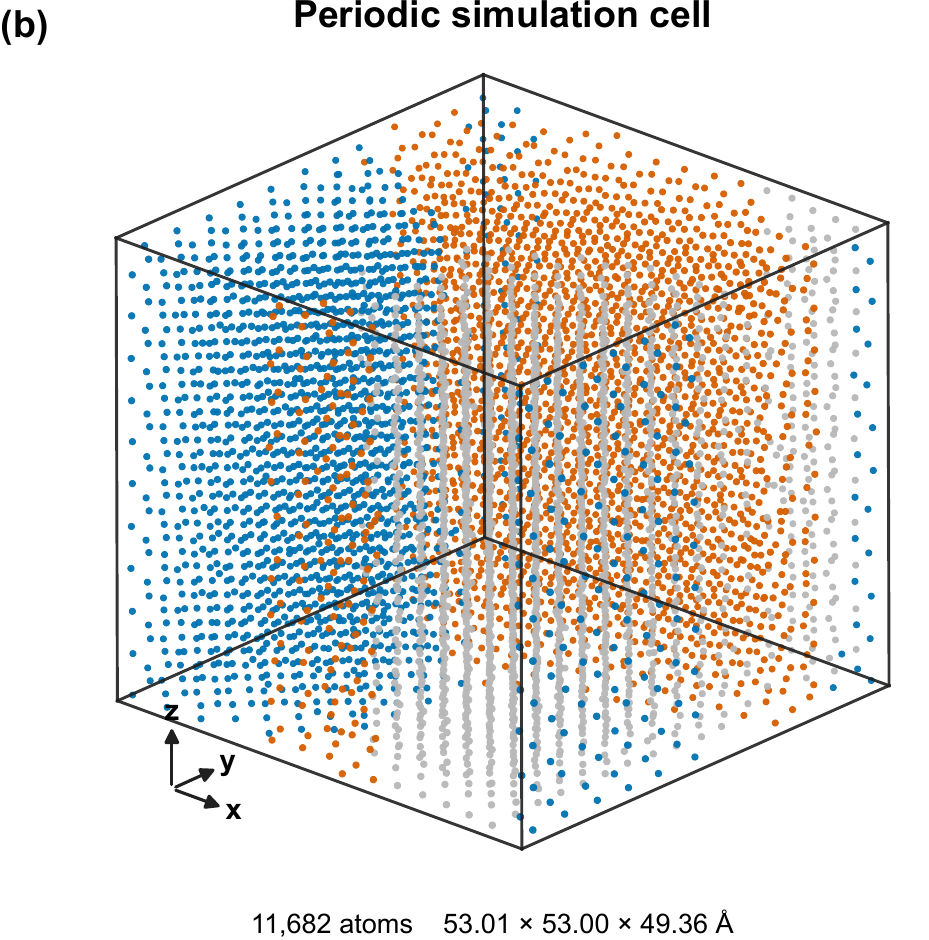}
        \phantomsubcaption
        \label{fig:hec_cell}
    \end{subfigure}
    \hfill
    \begin{subfigure}[b]{0.32\textwidth}
        \includegraphics[width=\linewidth]{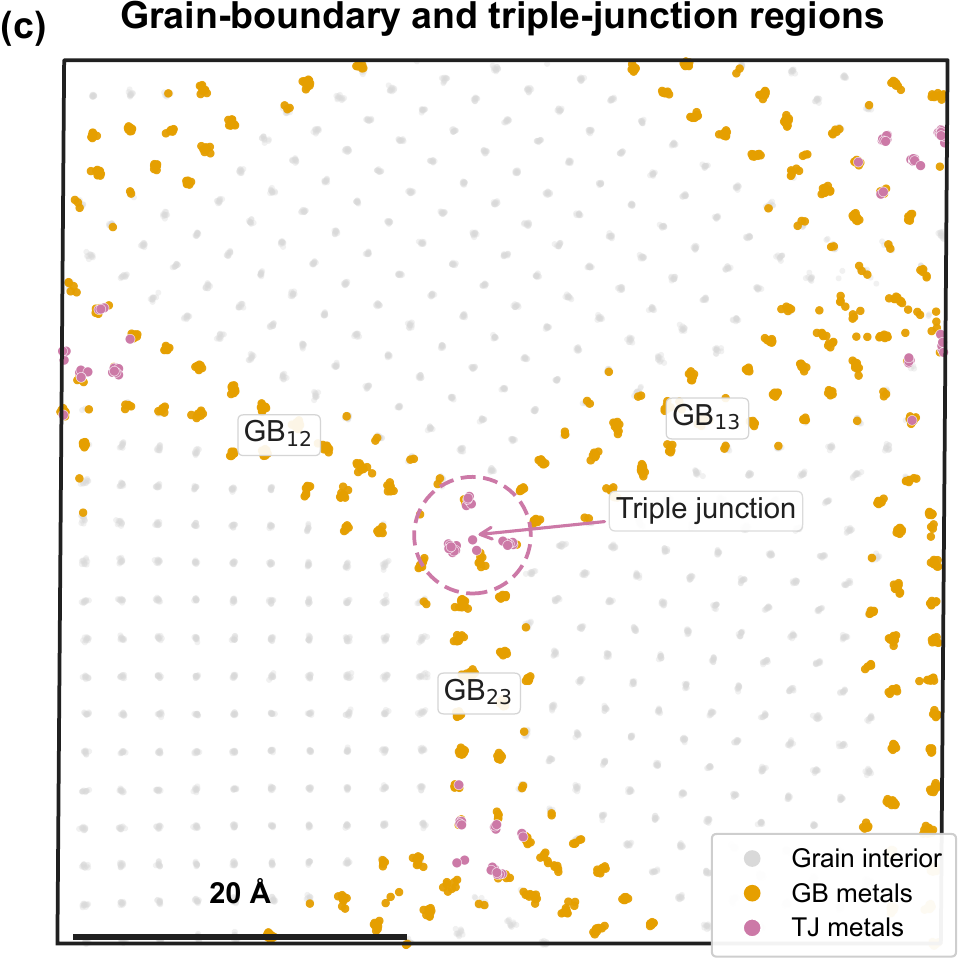}
        \phantomsubcaption
        \label{fig:hec_regions}
      \end{subfigure}
    }
  \caption{\justifying\small Three-grain polycrystalline HETMC model containing 11,682 atoms:
    {\bf (a)} grain morphology viewed along \(z\), {\bf (b)} periodic simulation cell, and
    {\bf (c)} grain-interior, grain-boundary, and triple-junction metals.
    Carbon atoms are omitted for clarity.}
  \label{fig:hec_polycrystal}
\end{figure*}

Despite segregation of the group-VI species in both compositions, the
Cr- and W-containing HETMCs exhibited markedly different fracture behavior.
The W-containing composition showed transgranular fracture and higher
compressive strength, whereas the Cr-containing composition showed
intergranular fracture, comparatively lower compressive strength, and pop-in events during indentation.
This contrast suggests that segregation alone does not determine the
GB response; the thermal stability and mechanical property of the segregated boundary
may also depend strongly on its chemistry.

The difference between W and Cr segregation may arise from a lower thermodynamic
driving force for W segregation, slower diffusion of W relative to Cr, or differences
in the state of the corresponding carbides during processing. For the latter possibility, the melting
temperature of $\mathrm{CrC}_{x}$ ranges from approximately 1550--1810$^\circ$C, depending on
stoichiometry, whereas WC melts above $2700^{\circ}\mathrm{C}$. Therefore,
$\mathrm{CrC}_{x}$ is likely to be liquid at the sintering temperature, while WC is
expected to remain solid.

The combined effects of segregation thermodynamics,
diffusion kinetics, and the physical state of the carbide at processing temperatures
raise the following questions: \\

\noindent {\bf i.} Is the concentration of interfacial Cr sufficient to induce premelting below the processing temperature? \\

\noindent {\bf ii.}  What role does the interfacial structure play in premelting, independent of overall composition? \\

\noindent {\bf iii.} $\mathrm{MoC}_{x}$ has a melting temperature at substoichiometric
    compositions that lies between those of $\mathrm{CrC}_{x}$ and
    $\mathrm{WC}_{x}$: is $\mathrm{MoC}_{x}$ therefore predicted to exhibit
    interfacial premelting below the processing temperature? \\

Recent universal machine-learning interatomic potentials (MLIPs) trained on broad inorganic-materials databases, such as OMat24~\cite{barrosoluque2026openmaterials2024omat24}, provide a practical framework for simulating complex multicomponent ceramics without composition-specific fitting~\cite{aflowML}. The MACE architecture~\cite{batatia_foundation_2025_etal}, in particular, has demonstrated transferable accuracy across diverse chemistries. This makes it practical to compare GB behavior across multiple HETMC compositions under identical simulation conditions, something impractical with composition-specific potentials or density functional theory.

\begin{figure*}[htb!]
    \includegraphics[width=0.99\textwidth]{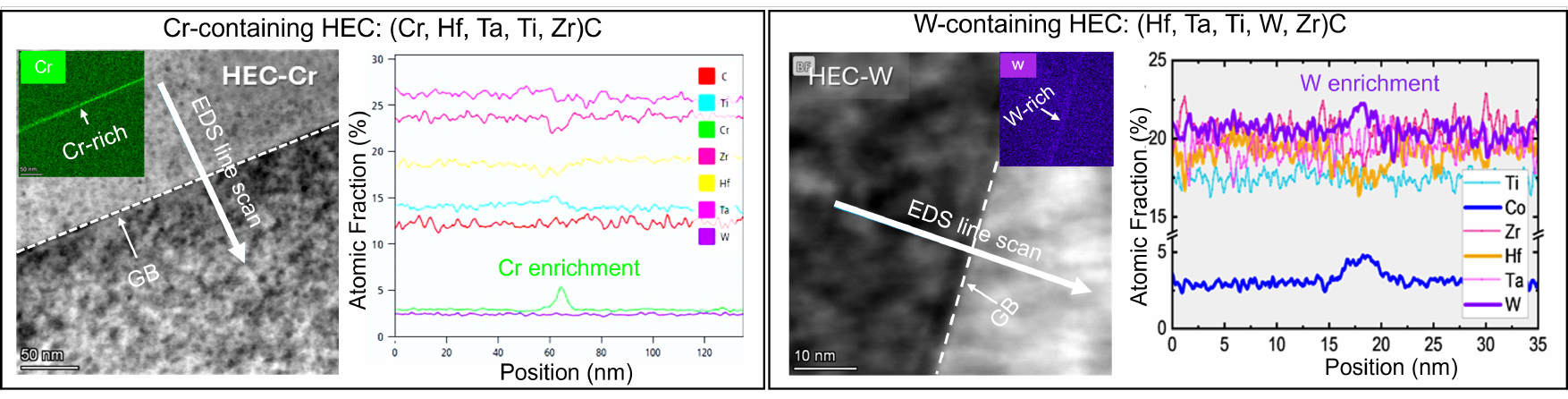}
    \caption{\justifying\small Electron-microscopy images and corresponding EDS
      line-scan profiles across grain boundaries
      in Cr- and W-containing high-entropy carbides. The left pair
      shows \ch{(Cr,Hf,Ta,Ti,Zr)C},
      while the right pair shows \ch{(Hf,Ta,Ti,W,Zr)C}. In each
      microscopy image, the dashed
      white line marks the GB, and the solid white
      arrow indicates the direction of the EDS line scan.
      The insets show the corresponding Cr and W elemental maps,
      revealing
      locally enriched regions near the GB.
      Adapted from Ref.~\onlinecite{schenck_effect_2026}.}
    \label{fig:exp_sim_cr_segregation}

\end{figure*}

To gain further insight into this interfacial melting behavior, we carried out
Monte Carlo (MC) and molecular dynamics (MD) simulations using the pretrained
universal MACE-OMAT-0 MLIP across five models representing four HETMC compositions. The Cr-containing compositions are \ch{(Cr,Hf,Ta,Ti,Zr)C} and \ch{(Cr,Nb,Ta,Ti,Zr)C}, while the Cr-free compositions are \ch{(Hf,Mo,Ta,Ti,Zr)C} and \ch{(Hf,Ta,Ti,W,Zr)C}. For \ch{(Cr,Hf,Ta,Ti,Zr)C}, both MC-segregated and chemically random GB configurations were considered. The simulated segregation trends were
compared with backscattered-electron (BSE) imaging and scanning transmission electron microscopy coupled with
energy-dispersive X-ray spectroscopy (STEM--EDS) observations of the Cr- and W-containing
materials, which showed Cr enrichment at the GBs and localized W-rich features along portions of the
GB network.
The simulations indicate preferential enrichment of Cr and Zr at the Cr-containing GBs and premelting-like disordering of the resulting Cr-rich interfaces below typical processing temperatures, consistent with conclusions drawn from prior experimental studies. In contrast, the Mo- and W-containing compositions exhibit delayed interfacial disordering, revealing a composition-dependent hierarchy of GB stability.

\section{Materials and Methods}

\subsection{Experimental}

The synthesis of \ch{(Cr,Hf,Ta,Ti,Zr)C} and
\ch{(Hf,Ta,Ti,W,Zr)C} specimens was reported previously~\cite{schenck_effect_2026}. Briefly,
nominally equimolar amounts of metal-oxide powders were combined with carbon black. The powders were mixed by high energy ball milling for 1~h, and then underwent carbothermally reduction at $1610^{\circ}\mathrm{C}$ for
3~h under mild vacuum. The resulting carbide powders were densified by
direct current sintering for 10~min under a uniaxial pressure of
$50\,\mathrm{MPa}$. The densification temperatures were
$2000^{\circ}\mathrm{C}$ for
\ch{(Cr,Hf,Ta,Ti,Zr)C} and
$1950^{\circ}\mathrm{C}$ for \ch{(Hf,Ta,Ti,W,Zr)C}. Polished sections were examined using a backscatter detector in a scanning electron microscopy (SEM; Hitachi SU3900) operated at $20\,\mathrm{kV}$. Chemical analysis was performed in-situ in the SEM using energy dispersive spectroscopy (EDS). Site-specific
transmission electron microscopy (TEM) lamellae were prepared by focused
ion beam (FIB) lift-out. Bright-field transmission electron microscopy
(BF-TEM) and STEM--EDS were performed at
$200\,\mathrm{kV}$ using Thermo Fisher Titan 80--300 and Talos F200X
microscopes. Composition profiles were extracted normal to selected GBs.

\begingroup
\setlength{\intextsep}{4pt}
\setlength{\abovecaptionskip}{2pt}
\setlength{\belowcaptionskip}{4pt}
\setlength{\tabcolsep}{4pt}
\renewcommand{\arraystretch}{1.02}

\sisetup{
    detect-all,
    table-number-alignment=center
}

\begin{table*}
\centering
\singlespacing
\small

\caption{\justifying Species-resolved grain-boundary segregation in four
HETMCs. Here, \(c_{\mathrm{GB}}\) and
\(c_{\mathrm{int}}\) denote the GB and grain-interior compositions,
respectively, and \(c_{\mathrm{GB}}/c_{\mathrm{ref}}\) is the enrichment
ratio relative to \(c_{\mathrm{ref}}=20.0\) at.\%.}
\label{tab:composition_compare_gb}

\begin{tabular*}{\textwidth}{
@{\extracolsep{\fill}}
l
S[table-format=2.1]
S[table-format=1.2]
S[table-format=2.1]
@{\hspace{1.2em}}
l
S[table-format=2.1]
S[table-format=1.2]
S[table-format=2.1]
}
\toprule

\multicolumn{4}{c}{Cr-containing carbides} &
\multicolumn{4}{c}{Cr-free carbides} \\

\cmidrule(lr){1-4}
\cmidrule(lr){5-8}

Species &
\multicolumn{1}{c}{\(c_{\mathrm{GB}}\)} &
\multicolumn{1}{c}{\(c_{\mathrm{GB}}/c_{\mathrm{ref}}\)} &
\multicolumn{1}{c}{\(c_{\mathrm{int}}\)} &
Species &
\multicolumn{1}{c}{\(c_{\mathrm{GB}}\)} &
\multicolumn{1}{c}{\(c_{\mathrm{GB}}/c_{\mathrm{ref}}\)} &
\multicolumn{1}{c}{\(c_{\mathrm{int}}\)} \\

&
\multicolumn{1}{c}{(at.\%)} &
&
\multicolumn{1}{c}{(at.\%)} &
&
\multicolumn{1}{c}{(at.\%)} &
&
\multicolumn{1}{c}{(at.\%)} \\

\midrule

\multicolumn{4}{l}{\ch{(Cr,Hf,Ta,Ti,Zr)C}} &
\multicolumn{4}{l}{\ch{(Hf,Mo,Ta,Ti,Zr)C}} \\

Cr & 45.6 & 2.28 & 12.8 &
Mo & 46.6 & 2.33 & 12.0 \\

Zr & 35.3 & 1.77 & 15.7 &
Zr & 33.3 & 1.66 & 16.0 \\

Hf & 5.6 & 0.28 & 24.1 &
Ti & 9.5 & 0.48 & 23.1 \\

Ta & 8.6 & 0.43 & 23.2 &
Ta & 5.9 & 0.29 & 24.2 \\

Ti & 4.8 & 0.24 & 24.3 &
Hf & 4.8 & 0.24 & 24.6 \\

\addlinespace[3pt]
\midrule
\addlinespace[2pt]

\multicolumn{4}{l}{\ch{(Cr,Nb,Ta,Ti,Zr)C}} &
\multicolumn{4}{l}{\ch{(Hf,Ta,Ti,W,Zr)C}} \\

Cr & 46.3 & 2.31 & 13.0 &
W  & 44.6 & 2.23 & 12.8 \\

Zr & 25.9 & 1.30 & 18.4 &
Zr & 32.3 & 1.62 & 16.4 \\

Nb & 19.5 & 0.97 & 20.1 &
Ti & 13.1 & 0.66 & 22.0 \\

Ta & 5.6 & 0.28 & 23.8 &
Hf & 5.7 & 0.28 & 24.2 \\

Ti & 2.7 & 0.13 & 24.6 &
Ta & 4.3 & 0.22 & 24.6 \\

\bottomrule
\end{tabular*}

\end{table*}
\endgroup

\subsection{Computational}
Polycrystalline models for all four compositions were generated by Voronoi
tessellation using Atomsk~\cite{hirel_atomsk_2015}. For all the models, a
\(53.0\times53.0\times4.5~\mathrm{\AA^3}\) rocksalt carbide slab was
partitioned into three grains using seed positions
\((9.4,14.6,2.3)\), \((39.5,15.6,2.3)\), and
\((23.9,40.6,2.3)~\mathrm{\AA}\), with corresponding in-plane
orientations of \(-178.9^\circ\), \(-136.7^\circ\), and
\(-33.5^\circ\). The initial 1062-atom slab was replicated 11 times along
the \(z\)-direction, producing a periodic
\(53.0\times53.0\times49.5~\mathrm{\AA^3}\) supercell containing
11,682 atoms. Representative views of the model are illustrated in
Figure~\ref{fig:hec_polycrystal}.

All MC energy evaluations and MD simulations were
performed using the pretrained MACE-OMAT-0 small potential without additional
training or fine-tuning~\cite{batatia_foundation_2025_etal}. MACE-OMAT-0 was selected because it enabled
large-scale finite-temperature sampling at substantially lower cost than DFT
and showed the best overall agreement with DFT among the universal MLIPs
tested in our previous study~\cite{Mou2026FiniteTemperature}. That validation reproduced the relative trends
in \(\Sigma5(210)\) carbide GB energies and entropy-forming ability (EFA), supporting
its use here for comparative analysis of GB disordering and single metal carbide
melting behavior. Before chemical sampling,
each structure was relaxed in LAMMPS~\cite{thompson_lammps_2022} by
conjugate-gradient minimization, followed by 100~fs of NVT equilibration and
500~fs of NPT equilibration at $27^\circ\mathrm{C}$~\cite{nose_unified_1984,hoover_canonical_1985,martyna_constant_1994, parrinello_crystal_1980,tuckerman_liouville-operator_2006}. Both equilibration stages used a 0.05~fs timestep to maintain stable integration of the as-constructed polycrystalline GB models, which can contain locally high forces after Voronoi construction and chemical randomization. The NPT stage was performed without external pressure and chemical ordering was sampled by MC atom swaps while keeping the
carbon sublattice fixed. The present MC sampling considers only metal redistribution in stoichiometric carbides and excludes carbon-vacancy effects on segregation. We followed the MC workflow described in our previous study~\cite{Mou2026FiniteTemperature}, except that no short MD segment was performed after each attempted swap and each calculation was extended to \(5\times10^{5}\) attempted swaps. At each step, two unlike metal atoms were selected randomly and their identities were exchanged while the carbon sublattice remained fixed. The trial energy was evaluated directly using the MACE-OMAT-0 potential, and the move was accepted according to the Metropolis criterion at \(300~\mathrm{K}\)~\cite{metropolis_equation_1953}. Sampling was continued until the potential energy reached a plateau.

The MC-sampled structures were subjected to two independent NPT heating
simulations at zero external pressure using the same MACE-OMAT-0 potential.
Both simulations used a timestep of 0.1~fs, thermostat and barostat damping
constants of 10 and 100~fs, respectively, with all cell dimension and shear degrees of freedom allowed to evolve independently. The first simulation was a broad-range anneal used to evaluate
species-resolved atomic mobility. Each structure was equilibrated at
approximately $27^\circ\mathrm{C}$ for 1.0~ps, heated from
$27^\circ\mathrm{C}$ to $1127^\circ\mathrm{C}$ over 2.2~ps, and then
heated to $2727^\circ\mathrm{C}$ over 16.0~ps, giving a total trajectory
length of 19.2~ps. This trajectory was used for the species-resolved
mean-squared-displacement (MSD) analysis. A separate, slower heating simulation was performed to resolve the onset of
GB disordering more accurately. The focused heating scan began with approximately $2.0~\mathrm{ps}$ of equilibration, followed by heating from $1127$ to $1927^\circ\mathrm{C}$ in four consecutive $200^\circ\mathrm{C}$ intervals at a constant rate of $25^\circ\mathrm{C\,ps^{-1}}$. Each interval lasted $8.0~\mathrm{ps}$, yielding approximately $32~\mathrm{ps}$ of analyzed heating data. The constant-temperature equilibration portions were excluded
from the analysis. These trajectories were used to calculate the local
Lindemann index and liquid-like pair fraction in the GB and grain-interior
regions.

\begin{figure*}[htb!]
    \includegraphics[width=0.99\textwidth]{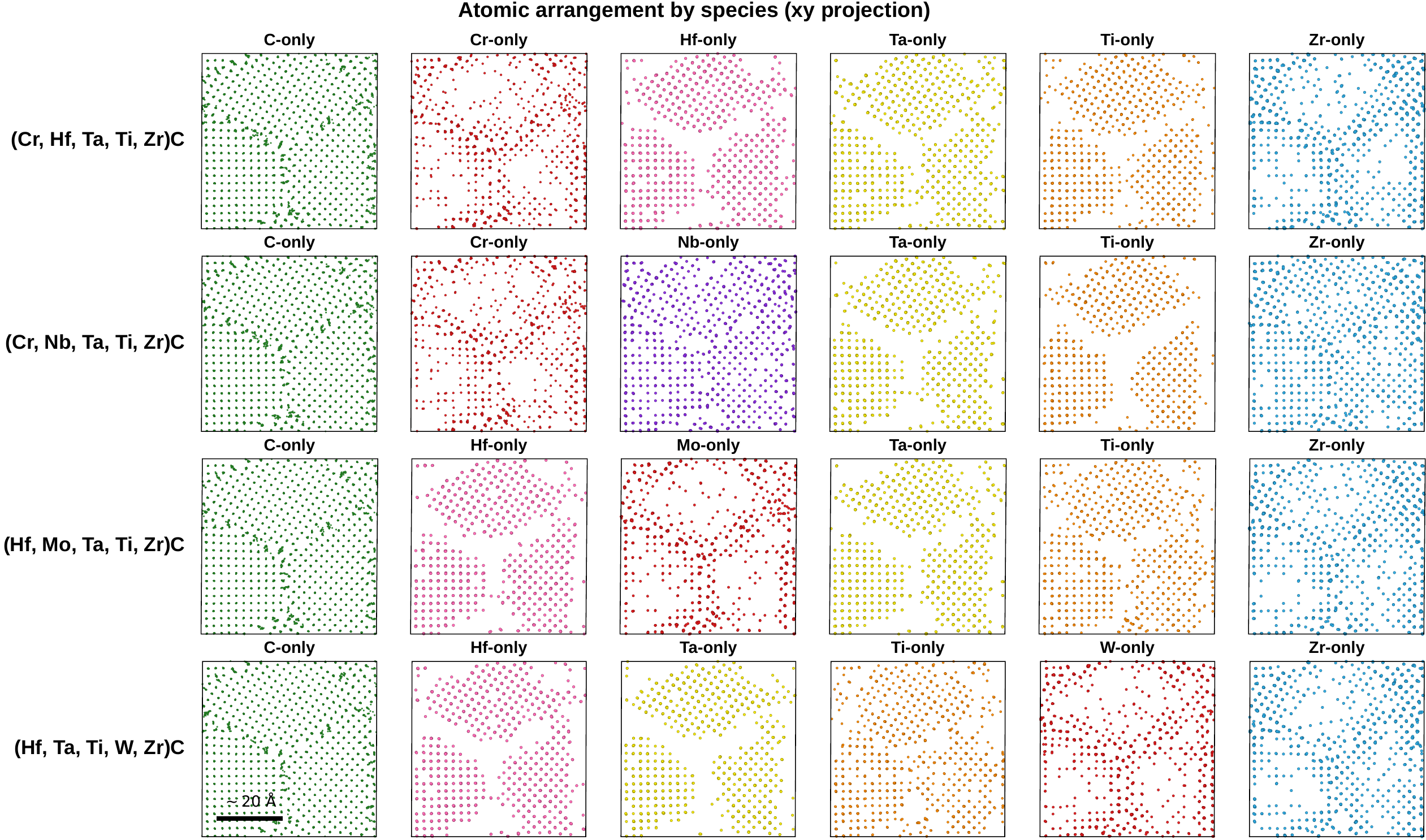}
\vspace{-2mm}
\caption{\justifying\small Element-resolved atomic projections of the polycrystalline high-entropy carbide models viewed along the [001] direction. From top to bottom, the rows correspond to \ch{(Cr,Hf,Ta,Ti,Zr)C}, \ch{(Cr,Nb,Ta,Ti,Zr)C}, \ch{(Hf,Mo,Ta,Ti,Zr)C}, and \ch{(Hf,Ta,Ti,W,Zr)C}. For each composition, the six panels show the spatial distributions of C and the five constituent metal species separately. The species colors are C (green), Cr, Mo, and W (red), Hf (pink), Nb (violet), Ta (yellow), Ti (orange), and Zr (blue). The scale bar in the bottom C species projection is approximately 20~\AA.}
    \label{fig:cr1_hec_species_map}
\end{figure*}

Premelting-like interfacial disorder was quantified from the slower,
focused heating trajectories spanning \(1127\)--\(1927^\circ\mathrm{C}\)
using a local pair-relative Lindemann index,
\(\delta_R\)~\cite{chakravarty_lindemann_2007}. The heating trajectories were divided into
\(50^{\circ}\mathrm{C}\) temperature intervals. The GB region and grain-interior regions were defined
relative to the initial three-grain geometry and retained throughout the
analysis. For a region \(R\),

\begin{equation}
\delta_R =
\frac{1}{N_{p,R}}
\sum_{(i,j)\in R}
\frac{
\sqrt{
\left\langle
\left|
\mathbf{r}_{ij}(t)-
\left\langle\mathbf{r}_{ij}(t)\right\rangle_t
\right|^2
\right\rangle_t
}
}{
\left\langle|\mathbf{r}_{ij}(t)|\right\rangle_t
},
\label{eq:lindemann}
\end{equation}

\noindent where \(\mathbf{r}_{ij}(t)\) is the minimum-image separation vector,
\(N_{p,R}\) is the number of local neighbor pairs in region \(R\), and
\(\langle\cdots\rangle_t\) denotes averaging over frames within a
temperature interval. C--C, C--metal, and metal--metal pairs were included.
The individual term within the summation defines the pair-resolved
Lindemann index, \(\delta_{ij}\).

A threshold of \(\delta_{ij}=0.15\) was used as an operational indicator of
substantial local disorder~\cite{chakravarty_lindemann_2007}. The
liquid-like pair fraction in region \(R\) was calculated as

\begin{equation}
f_{\mathrm{liq},R}
=
\frac{
N_{p,R}\!\left(\delta_{ij}\geq 0.15\right)
}{
N_{p,R}
},
\label{eq:liquid_fraction}
\end{equation}

\noindent where \(N_{p,R}(\delta_{ij}\geq0.15)\) is the number of local pairs whose
pair-resolved Lindemann index reaches or exceeds 0.15. This quantity
represents the fraction of locally disordered pairs and was not interpreted
as a thermodynamic liquid volume fraction. Because the critical Lindemann
value depends on the material, bonding environment, and definition of
\(\delta\), the threshold was not treated as absolute evidence of a fully
liquid state~\cite{jiang_thermal_2007}. Instead, disordering onset was
identified by comparing the GB region and grain-interior responses.
Threshold-crossing temperatures were estimated from the smoothed
\(\delta_R\) curves using linear interpolation between adjacent temperature
intervals. Smoothing was applied for visualization and estimation of
the crossing temperatures.

MSDs were evaluated for C and the
relevant segregating metal species, including Cr, Zr, Mo, W, and Nb. GB
atoms were selected within \(2.5~\text{\AA}\) of the reference boundary,
whereas grain-interior atoms were selected from the farthest \(30\%\) of
each species. Equal-sized GB and grain-interior populations were used
separately for each species and composition to minimize sampling bias.
The MSD values were binned in \(50^\circ\mathrm{C}\) intervals and
smoothed for visualization.

Single metal carbide melting behavior was evaluated separately using one-phase
NPT heating simulations of initially rocksalt carbide structures. Each
structure was first minimized with isotropic cell relaxation at zero
external pressure, initialized at approximately \(27^\circ\mathrm{C}\),
and continuously heated to \(4227^\circ\mathrm{C}\) at
\(10^\circ\mathrm{C\,ps^{-1}}\). The \(420~\mathrm{ps}\) ramp comprised
\(4.2\times10^{5}\) MD steps using a \(1.0~\mathrm{fs}\) timestep, with
thermostat and barostat damping constants of \(0.1\) and
\(1.0~\mathrm{ps}\), respectively. Enthalpy, volume, density, and MSD were
sampled every \(1~\mathrm{ps}\), while atomic configurations were stored
every \(5~\mathrm{ps}\). Transition temperatures were identified from simultaneous increases in enthalpy, volume, and MSD. A composite score combining these changes was used to identify
the dominant transition region, and the score-weighted average temperature within
this region was taken as the one-phase heating melting indicator. Because the simulations involved continuous heating of a single crystalline phase, these temperatures were interpreted as one-phase heating melting indicators rather than equilibrium melting points, since crystalline superheating may occur~\cite{zhang2012comparison}.

\begin{figure*}
    \includegraphics[width=0.99\textwidth]{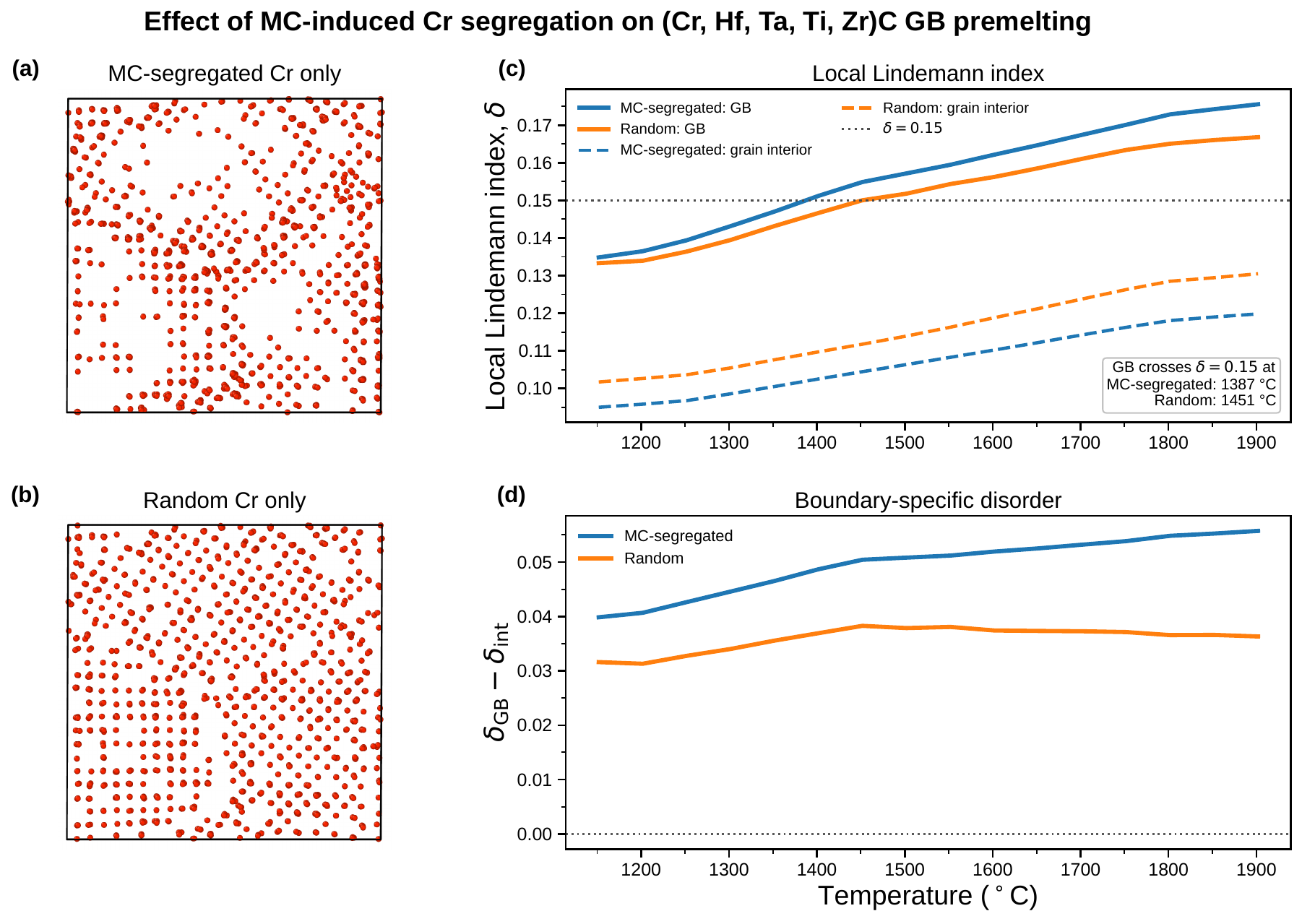}
    \vspace{-2mm}
    \caption{\justifying\small Effect of MC-induced chemical ordering and Cr segregation on
    grain-boundary premelting in
    \ch{(Cr,Hf,Ta,Ti,Zr)C}.
    {\bf (a,b)} Cr-only atomic projections of the MC-segregated and random
    configurations, respectively.
    {\bf (c)} Local Lindemann indices of the GB core and grain interior; the dotted
    line marks \(\delta=0.15\).
    {\bf (d)} Boundary-specific excess disorder,
    \(\delta_{\mathrm{GB}}-\delta_{\mathrm{int}}\).}
    \label{fig:Cr_ordered_random_lindemann}
\end{figure*}

\section{Results and discussion }

\subsection{Experimental observations of grain-boundary segregation}

Electron microscopy and STEM--EDS line-scan measurements showed composition-dependent elemental enrichment at selected GBs in the Cr- and W-containing HETMCs (Figure~\ref{fig:exp_sim_cr_segregation}).
In \ch{(Cr,Hf,Ta,Ti,Zr)C}, the Cr elemental map shows a Cr-rich region associated with the analyzed boundary.
The corresponding EDS profile exhibits a distinct, localized increase in the Cr atomic fraction near the position at which the scan crosses the GB, while the concentrations of the other measured elements remain comparatively uniform.
This observation shows that Cr enrichment occurs locally along individual grain boundaries and is not restricted to larger Cr-rich features observed elsewhere in the microstructure.

In \ch{(Hf,Ta,Ti,W,Zr)C}, the W elemental map similarly shows a localized W-rich region near the analyzed boundary.
The corresponding line profile exhibits a modest but distinct increase in the W atomic fraction near the GB position. Compared with the pronounced Cr peak, the W enrichment in this selected region is weaker and distributed over a somewhat broader distance.
These measurements qualitatively support the simulated tendency of both Cr and W to segregate toward interfacial regions.

Additional microscopy reported previously showed Cr-rich regions at GBs and triple junctions, as well as W-rich platelets or nanograins along selected portions of the grain-boundary network.
High-resolution STEM and diffraction further identified the latter features as a W-rich phase, with local carbon depletion observed in some regions.
Detailed synthesis, microscopy, diffraction, and compositional analyses were reported previously~\cite{schenck_effect_2026}.

\subsection{Computational grain-boundary segregation and experimental comparison}
The final MC-sampled structures reveal pronounced chemical heterogeneity
across all four polycrystalline HETMC compositions. Table~\ref{tab:composition_compare_gb}
summarizes the corresponding GB segregation behavior, where
\(c_{\mathrm{ref}}\) denotes the overall metal composition,
\(c_{\mathrm{GB}}\) the GB composition, and \(c_{\mathrm{interior}}\) the
grain-interior composition, while \(c_{\mathrm{GB}}/c_{\mathrm{ref}}\)
quantifies enrichment or depletion relative to the equimolar reference
state.

In each composition, the group-VI element is the most strongly
enriched GB species. Cr reaches 45.6~at.\% in
\ch{(Cr,Hf,Ta,Ti,Zr)C} and 46.3~at.\% in
\ch{(Cr,Nb,Ta,Ti,Zr)C}, corresponding to enrichment ratios
of 2.28 and 2.31, respectively. Similarly, Mo reaches 46.6~at.\%
(\(c_{\mathrm{GB}}/c_{\mathrm{ref}}=2.33\)) in
\ch{(Hf,Mo,Ta,Ti,Zr)C}, while W reaches 44.6~at.\%
(\(c_{\mathrm{GB}}/c_{\mathrm{ref}}=2.23\)) in
\ch{(Hf,Ta,Ti,W,Zr)C}. Zr is also enriched in all four
systems, with GB concentrations ranging from 25.9 to 35.3~at.\%. The strong
Cr enrichment is consistent with experimental STEM--EDS observations of Cr
segregation in \ch{(Cr,Hf,Ta,Ti,Zr)C}
\cite{schenck_effect_2026} and \ch{(Cr,Nb,Ta,Ti,Zr)C}
\cite{su_fracture_2024}. In contrast, Ta and Hf are strongly depleted at
the GBs, Ti is depleted to varying degrees, and Nb remains nearly neutral in
the Nb-containing composition. These quantitative trends are consistent
with the element-resolved projections in
Figure~\ref{fig:cr1_hec_species_map}, which show non-random cation
redistribution along the GB network while the overall carbide lattice
remains intact. At the 300~K sampling temperature, these similar concentrations reflect a common preference for boundary enrichment in the low-energy configurations sampled here, while the different Cr and W concentrations observed experimentally likely reflect additional processing-dependent effects. Two caveats to this are that the number of each species is kept constant while they can be lost through flow or other means in the experiment, and we have not explored the possible role of carbon vacancies.

\subsection{Composition-dependent grain-boundary disordering}
\begin{figure*}
\includegraphics[width=0.99\textwidth]{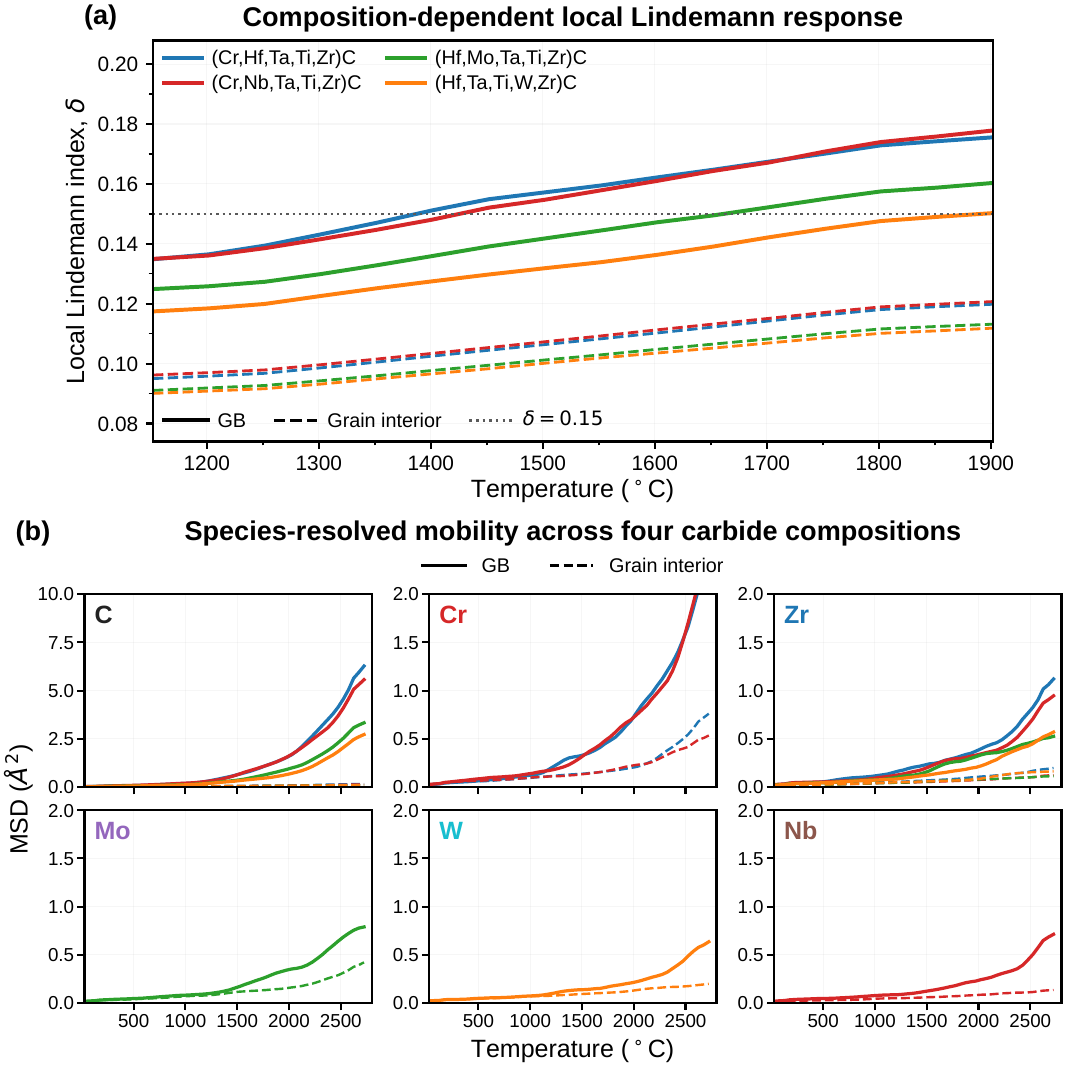}
    \vspace{-3mm}
    \caption{\justifying\small Comparative thermal disordering and atomic mobility across four HETMCs.
      {\bf (a)} Composition-dependent local Lindemann response for the grain-boundary region and grain interior. Solid and dashed curves represent the GB and grain-interior responses, respectively, and the dotted line marks \(\delta = 0.15\).
    {\bf (b)} Species-resolved MSD for C, Cr, Zr, Mo, W, and Nb in the GB region and grain interior during annealing from approximately \(27\) to \(2727^{\circ}\mathrm{C}\).}
    \label{fig:lindemann_msd_combined}
\end{figure*}

Because MC sampling generates chemically heterogeneous GB
regions, their thermal stability was evaluated using the focused heating
simulations. Figure~\ref{fig:lindemann_msd_combined}(a) compares the
all-atom local Lindemann indices of the GB and grain-interior regions across the four HETMC compositions. The
chemically random and MC-segregated
\ch{(Cr,Hf,Ta,Ti,Zr)C} configurations are compared separately in
Figure~\ref{fig:Cr_ordered_random_lindemann}. A clear hierarchy of
interfacial disordering is observed. The MC-segregated
\ch{(Cr,Hf,Ta,Ti,Zr)C} model reaches the adopted threshold at
\(T_{\mathrm{cross}}\approx1390^{\circ}\mathrm{C}\), followed by
\ch{(Cr,Nb,Ta,Ti,Zr)C} at approximately
\(1430^{\circ}\mathrm{C}\) and chemically random
\ch{(Cr,Hf,Ta,Ti,Zr)C} at approximately
\(1450^{\circ}\mathrm{C}\).

The effect of MC-induced chemical ordering is examined more directly in
Figure~\ref{fig:Cr_ordered_random_lindemann}. The Cr-only projections show
pronounced Cr accumulation along the GB network in the MC-segregated structure,
whereas the random configuration retains a more uniform Cr distribution
(Figure~\ref{fig:Cr_ordered_random_lindemann}(a,b)). The corresponding
Lindemann responses confirm earlier and more strongly localized disordering
in the MC-segregated configuration
(Figure~\ref{fig:Cr_ordered_random_lindemann}(c,d)). The boundary-specific excess disorder,
\(\delta_{\mathrm{GB}}-\delta_{\mathrm{int}}\), increases from approximately
\(0.040\) to \(0.056\) for the MC-segregated structure, whereas the random
configuration remains between approximately \(0.031\) and \(0.038\). The approximately \(60^{\circ}\mathrm{C}\) lower threshold-crossing temperature, along with the larger boundary-specific excess disorder, indicates that MC-induced chemical ordering advances the onset of GB disorder and makes it more localized. Because MC
sampling redistributes all metal species, this behavior is attributed to the
resulting Cr-rich interfacial chemistry rather than to Cr alone.

The two Cr-free models exhibit substantially delayed GB disordering.
The \ch{(Hf,Mo,Ta,Ti,Zr)C} GB reaches $\delta=0.15$ near
$1660^{\circ}\mathrm{C}$, whereas the
\ch{(Hf,Ta,Ti,W,Zr)C} GB does not reach the threshold until
approximately $1890^{\circ}\mathrm{C}$. In all five models, the
grain-interior Lindemann indices remain below 0.15 throughout the analyzed
temperature range, showing that substantial disorder develops
preferentially at the interfaces rather than uniformly throughout the
grains.  Because
the MD heating rate is much faster than experimental heating rates, these
temperatures should be viewed as comparative simulation metrics; slower heating
or experiments may shift the observed onset.

\subsection{Liquid-like pair fraction and species-resolved atomic mobility}

To complement the mean Lindemann index, the liquid-like pair fraction,
\(f_{\mathrm{liq}}=f_{\delta_{ij}\geq0.15}\), was calculated for the GB and
grain-interior regions. This metric represents the fraction of local atomic
pairs whose pair-resolved displacement fluctuations exceed the
Lindemann threshold and therefore indicates how broadly strong local
disordering is distributed within each region. Table~\ref{tab:gb_bulk_disordering}
summarizes the corresponding GB threshold-crossing temperatures and the
disordering metrics evaluated near \(1900^{\circ}\mathrm{C}\) for the five
simulated models spanning four HETMC compositions.
\begin{table}[!t]
\centering
\scriptsize
\setlength{\tabcolsep}{2.0pt}
\renewcommand{\arraystretch}{1.10}

\caption{\justifying Comparison of GB and grain-interior disordering across the five
HETMC models. \(T_{\mathrm{cross}}\) is the temperature at which the GB
Lindemann index first reaches the liquid-like threshold,
\(\delta_{\mathrm{GB}}=0.15\). The GB and interior Lindemann indices,
liquid-like GB pair fraction \(f_{\mathrm{liq,GB}}\), and GB-to-interior
Lindemann-index ratio are reported at \(1900^{\circ}\mathrm{C}\).}
\label{tab:gb_bulk_disordering}

\resizebox{\columnwidth}{!}{
\begin{tabular}{@{}lccccc@{}}
\toprule
Composition &
\shortstack{\(T_{\mathrm{cross}}\)\\(\(^{\circ}\mathrm{C}\))} &
\(\delta_{\mathrm{GB}}\) &
\(\delta_{\mathrm{int}}\) &
\(f_{\mathrm{liq,GB}}\) &
\(\delta_{\mathrm{GB}}/\delta_{\mathrm{int}}\) \\
\midrule

\shortstack[l]{MC-segregated\\
\ch{(Cr,Hf,Ta,Ti,Zr)C}}
& 1390 & 0.176 & 0.120 & 0.51 & 1.47 \\

\shortstack[l]{MC-segregated\\
\ch{(Cr,Nb,Ta,Ti,Zr)C}}
& 1430 & 0.178 & 0.121 & 0.52 & 1.47 \\

\shortstack[l]{Chemically random\\
\ch{(Cr,Hf,Ta,Ti,Zr)C}}
& 1450 & 0.167 & 0.130 & 0.47 & 1.28 \\

\shortstack[l]{MC-segregated\\
\ch{(Hf,Mo,Ta,Ti,Zr)C}}
& 1660 & 0.160 & 0.113 & 0.41 & 1.42 \\

\shortstack[l]{MC-segregated\\
\ch{(Hf,Ta,Ti,W,Zr)C}}
& 1890 & 0.150 & 0.112 & 0.35 & 1.34 \\

\bottomrule
\end{tabular}
}
\end{table}

At approximately \(1900^{\circ}\mathrm{C}\), the MC-segregated
\ch{(Cr,Hf,Ta,Ti,Zr)C} and
\ch{(Cr,Nb,Ta,Ti,Zr)C} boundaries exhibit the largest
liquid-like pair fractions, with
\(f_{\mathrm{liq,GB}}\approx0.52\). Thus, more than half of the local
neighbor pairs within these Cr-rich GB regions exceed the adopted disordering
threshold. The chemically random
\ch{(Cr,Hf,Ta,Ti,Zr)C} boundary reaches a slightly lower value of
approximately 0.47, whereas the Mo- and W-containing boundaries reach
approximately 0.41 and 0.35, respectively. This comparison indicates that strong local disorder is most extensive
in the MC-segregated Cr-containing boundaries and is progressively reduced in
the more refractory Mo- and W-containing systems.

\begin{figure*}
    \centering

\includegraphics[width=\textwidth]{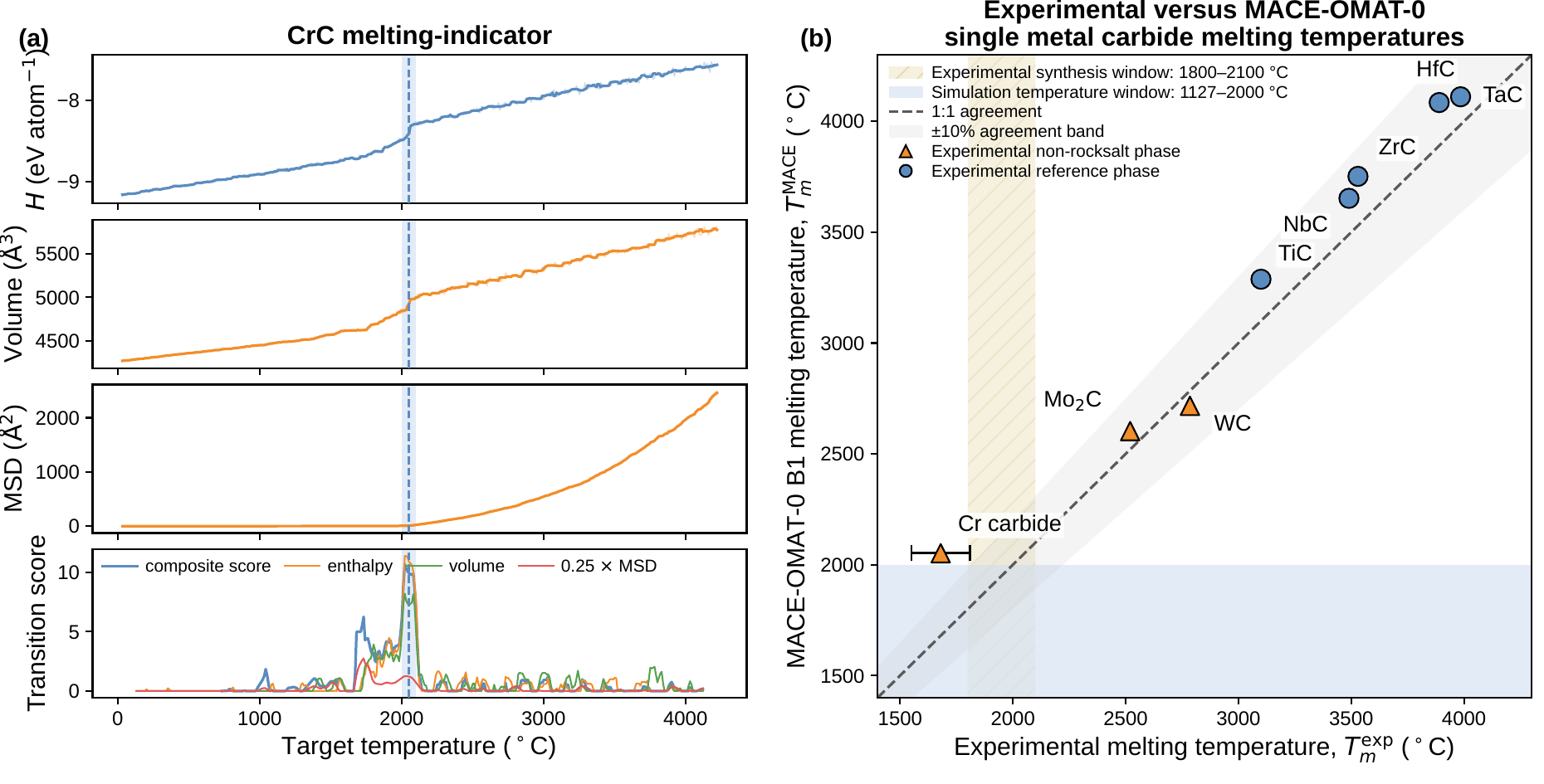}

    \vspace{-2mm}

    \caption{\justifying\small
    {\bf (a)} CrC one-phase heating melting indicator determined from changes in
    enthalpy, volume, MSD, and the transition score. The shaded region marks the
    transition interval and the dashed line its midpoint.
    {\bf (b)} Comparison of experimental melting temperatures with MACE-OMAT-0
    predictions for single metal carbides. The dashed line and gray
    band indicate \(1{:}1\) agreement and a \(\pm10\%\) range, respectively.
    The yellow and blue shaded regions mark the experimental synthesis window
    (\(1800\)--\(2100^{\circ}\mathrm{C}\)) and simulation window for Lindemann analysis
    (\(1127\)--\(2000^{\circ}\mathrm{C}\)). Blue circles and orange triangles
    denote experimental rocksalt and non-rocksalt reference phases, respectively;
    the horizontal error bar shows the reported Cr-carbide melting range.}

    \label{fig:binary_carbide_melting}
\end{figure*}

The GB-to-interior Lindemann-index ratio provides an additional measure of
the spatial localization of disorder. At \(1900^{\circ}\mathrm{C}\), the
two MC-segregated Cr-containing models exhibit ratios of approximately 1.47,
indicating nearly 50\% greater disorder at the GB than in the grain
interior. In contrast, the chemically random
\ch{(Cr,Hf,Ta,Ti,Zr)C} model has the smallest ratio, 1.28,
because its higher grain-interior Lindemann index reduces the contrast
between the two regions. The Mo- and W-containing models show intermediate
ratios of 1.42 and 1.34, respectively.

\begingroup
\setlength{\tabcolsep}{4pt}
\renewcommand{\arraystretch}{1.05}

\begin{table*}
\centering
\singlespacing
\small

\caption{\justifying MACE-OMAT-0-predicted single metal carbide one-phase
heating melting indicators. \(\Delta H\) and \(\Delta V\) were calculated
across the identified transition interval. MSD slopes were evaluated
immediately before and after the transition.}
\label{tab:binary_carbide_melting}

\begin{tabular*}{\textwidth}{
@{\extracolsep{\fill}}
l
c
c
S[table-format=1.3]
S[table-format=2.2]
S[table-format=1.3]
S[table-format=2.3]
}
\toprule

Carbide &
\multicolumn{1}{c}{Melting indicator} &
\multicolumn{1}{c}{Transition interval} &
\multicolumn{1}{c}{\(\Delta H\)} &
\multicolumn{1}{c}{\(\Delta V\)} &
\multicolumn{1}{c}{MSD slope before} &
\multicolumn{1}{c}{MSD slope after} \\

&
\multicolumn{1}{c}{(\(^{\circ}\mathrm{C}\))} &
\multicolumn{1}{c}{(\(^{\circ}\mathrm{C}\))} &
\multicolumn{1}{c}{(eV/atom)} &
\multicolumn{1}{c}{(\%)} &
\multicolumn{1}{c}{(\(\mathrm{\AA^2\,ps^{-1}}\))} &
\multicolumn{1}{c}{(\(\mathrm{\AA^2\,ps^{-1}}\))} \\

\midrule

CrC & 2052 & 2003--2103 & 0.195 & 3.14  & 0.643 & 2.965 \\
MoC & 2603 & 2533--2663 & 0.307 & 6.66  & 0.649 & 4.668 \\
WC  & 2717 & 2663--2773 & 0.255 & 6.50  & 0.422 & 3.396 \\
TiC & 3288 & 3233--3343 & 0.465 & 9.43  & 0.920 & 8.445 \\
NbC & 3653 & 3583--3713 & 0.539 & 10.24 & 1.024 & 10.567 \\
ZrC & 3752 & 3673--3823 & 0.606 & 12.99 & 0.765 & 8.956 \\
HfC & 4085 & 4033--4123 & 0.194 & 3.19  & 0.300 & 0.842 \\
TaC & 4111 & 4093--4123 & 0.101 & 1.28  & 0.229 & 0.343 \\

\bottomrule
\end{tabular*}

\end{table*}
\endgroup

The species-resolved MSD curves in
Figure~\ref{fig:lindemann_msd_combined}(b) show which species become mobile during the broad-range annealing simulation.
Because this simulation used a different temperature ranges and heating
rate from the focused Lindemann calculations, the MSD results are used to
identify relative mobility trends rather than to determine
threshold-crossing temperatures. The enhanced GB MSD provides complementary evidence for
premelting-like behavior, indicating increasing atomic mobility
within the disordered interfacial region~\cite{gao_universal_2026}. For the relevant constituent species, the
GB populations generally exhibit larger displacements than the corresponding
grain-interior populations. Carbon shows the strongest mobility response,
while the segregating group-VI elements and Zr become increasingly mobile at
elevated temperature. C atoms in the GB undergo a marked increase in displacement, whereas
grain-interior C atoms remain largely localized. This behavior is consistent
with vacancy-mediated carbon transport in transition-metal carbides, which can
be substantially faster than metal-sublattice diffusion
\cite{khatri_first-principles_2025,salehin_vacancy-cluster_2021}. Although C vacancies were not intentionally
introduced in the present models, local
undercoordination associated with the polycrystalline GB structure may
also contribute to the enhanced interfacial C mobility. Cr and Zr
show lower absolute MSD values but become increasingly mobile at elevated
temperature, indicating that early carbon-sublattice disorder is followed by
metal-species rearrangement within the chemically heterogeneous GB region. The
enhanced Cr mobility is also consistent with reports of Cr-assisted
densification and diffusion during high-temperature processing of HETMCs
\cite{su_fracture_2024}.

\subsection{Single metal carbide melting-point trends and implications for
grain-boundary stability}

The GB disordering sequence broadly follows the
melting-point hierarchy of the corresponding single metal carbides. CrC is shown as a representative
example in Figure~\ref{fig:binary_carbide_melting}(a). Coincident increases
in enthalpy and volume, together with the onset of rapid MSD growth, define
a transition region from approximately \(2003\) to \(2103^{\circ}\mathrm{C}\).
The transition score identifies the dominant change within this region,
giving a one-phase heating melting indicator of
\(2052^{\circ}\mathrm{C}\).

The MACE-OMAT-0 one-phase heating melting indicators for all eight single metal carbides
are summarized in Table~\ref{tab:binary_carbide_melting}. As shown in
Figure~\ref{fig:binary_carbide_melting}(b), the MACE-OMAT-0 predictions preserve
the overall experimental ranking, from the comparatively low melting
temperature of the Cr--C system to the highly refractory HfC and TaC end
members. The predicted melting temperatures are
\(2052^{\circ}\mathrm{C}\) for B1 CrC,
\(2603^{\circ}\mathrm{C}\) for B1 MoC,
\(2717^{\circ}\mathrm{C}\) for B1 WC,
\(3288^{\circ}\mathrm{C}\) for TiC,
\(3653^{\circ}\mathrm{C}\) for NbC,
\(3752^{\circ}\mathrm{C}\) for ZrC,
\(4085^{\circ}\mathrm{C}\) for HfC, and
\(4111^{\circ}\mathrm{C}\) for TaC. These values broadly track the
experimental melting-point trend. However, the comparisons for Cr,  and Mo
are not phase- or stoichiometry-matched because the calculations used 1:1 B1
structures, whereas the experimental references correspond to the commonly
reported stable carbide phases. For W, the WC stoichiometry is matched, but the calculated cubic B1 structure differs from the experimentally stable hexagonal WC structure. The agreement should therefore be interpreted
as a qualitative trend rather than a direct validation of the individual
melting temperatures.

This trend is consistent with the earlier disordering observed in the
Cr-enriched GBs. The first Lindemann-index crossings near
\(1390\)--\(1450^{\circ}\mathrm{C}\) occur below both the experimental
chromium-carbide melting range of approximately
\(1550\)--\(1810^{\circ}\mathrm{C}\) and the predicted melting temperature
of B1 CrC. These crossings therefore indicate the onset of substantial local
interfacial disorder rather than complete GB melting. With increasing
temperature, the simultaneous rise in the GB Lindemann index, liquid-like pair
fraction, and atomic mobility shows that disorder becomes progressively more
extensive within the Cr-rich interfaces. This behavior may contribute to the
experimentally observed redistribution of Cr-rich material along GBs during high-temperature consolidation~\cite{wang_role_2022,su_insights_2023,su_fracture_2024,
schenck_effect_2026}.

By comparison, the Mo- and W-containing GBs remain more ordered over the same
temperature range, consistent with the higher melting temperatures of the
corresponding single metal carbides. Near \(1900^{\circ}\mathrm{C}\), the Mo- and
W-containing GBs reach
\(f_{\mathrm{liq,GB}}\approx0.41\) and \(0.35\), respectively, and neither
reaches the half-disordered condition within the simulated window. The other
constituent carbides are also substantially more refractory, with experimental
melting temperatures of approximately
\(3100^{\circ}\mathrm{C}\) for TiC,
\(3490^{\circ}\mathrm{C}\) for NbC,
\(3530^{\circ}\mathrm{C}\) for ZrC,
\(3890^{\circ}\mathrm{C}\) for HfC, and
\(3985^{\circ}\mathrm{C}\) for TaC
~\cite{pierson1996handbook}. Their high melting temperatures are consistent
with the greater structural stability retained in the grain interiors.

Because HETMC consolidation commonly occurs near
\(1800\)--\(2100^{\circ}\mathrm{C}\), all GBs reach the adopted
disordering threshold within or below the processing window. However,
the Cr-rich interfaces disorder earlier and more extensively, whereas
the Mo- and especially W-containing GBs retain comparatively greater
structural order at processing temperatures. This
interpretation remains qualitative because the GB response also depends on
local chemistry, interfacial structure, heating rate, and simulation
timescale.

\section{Conclusions}

A comparative study of four high-entropy transition metal carbides  using Monte Carlo sampling and molecular dynamics with the
pretrained universal MACE-OMAT-0 interatomic potential shows a clear composition-dependent
trend in grain-boundary premelting-like disordering, largely controlled by
the chemistry of the segregated boundary region. The key findings are the following. \\

\noindent {\bf I.} MC sampling produced strongly non-random grain-boundary chemistries. Cr was the
dominant segregating species in the Cr-containing systems, while Mo and W
dominated the corresponding Cr-free boundaries. Zr was also enriched in all
four compositions. \\

\noindent {\bf II.} Experimental BSE and STEM--EDS observations qualitatively supported
the simulated segregation trends. Cr enrichment was observed at grain
boundaries in
\ch{(Cr,Hf,Ta,Ti,Zr)C}, whereas localized W-rich
interfacial features were identified along portions of the GB network in
\ch{(Hf,Ta,Ti,W,Zr)C}. \\

\noindent {\bf III.} The GB-disordering onset temperature increased in the order
Cr-containing $<$ Mo-containing $<$ W-containing. The MC-segregated
\ch{(Cr,Hf,Ta,Ti,Zr)C} boundary exceeded the Lindemann
threshold approximately $1390^{\circ}\mathrm{C}$, while
\ch{(Cr,Nb,Ta,Ti,Zr)C} crossed near
$1430^{\circ}\mathrm{C}$. Replacing Hf with Nb therefore had little effect
on the initial response of the Cr-rich boundary. \\

\noindent {\bf IV.} The chemically random
\ch{(Cr,Hf,Ta,Ti,Zr)C} model crossed near
$1450^{\circ}\mathrm{C}$. Thus, concentrating Cr at the grain boundary advanced the
resolved disordering onset by at least approximately
$60^{\circ}\mathrm{C}$ and increased the contrast between the GB and grain
interior. \\

\noindent {\bf V.} The Cr-free boundaries were more thermally stable. The Mo- and
W-containing compositions crossed the threshold near
$1660^{\circ}\mathrm{C}$ and $1890^{\circ}\mathrm{C}$, respectively,
while the grain interiors remained below the adopted threshold throughout
the focused heating range. \\

\noindent {\bf VI.} Near $1900^{\circ}\mathrm{C}$, the MC-segregated Cr-rich boundaries
showed the greatest extent of liquid-like disorder, followed by the
chemically random Cr-containing, Mo-containing, and W-containing systems.
The MSD results likewise showed that atomic mobility remained localized at
the grain boundary, with carbon exhibiting the strongest response. \\

\noindent {\bf VII.} The disordering hierarchy correlates qualitatively with the relative
refractoriness of the carbides concentrated at the boundaries. Cr-rich
interfaces may facilitate interfacial transport and densification, whereas
Mo- and particularly W-enriched boundaries provide greater resistance to
high-temperature disordering.

\section*{Acknowledgements}

Funding for this research was provided by the Office of Naval
Research through a Multidisciplinary University Research Initiative program
under project nos. N00014-21-1-2768, N00014-21-1-2515 and N00014-24-1-2768.
The authors thank Tarek Haque, Dr. Simon Divilov and Dr. Xiomara Campilongo for fruitful discussions.

\section*{Declaration of Generative AI and AI-assisted technologies in the manuscript preparation process}

During the preparation of this work, the author(s) used ChatGPT to rephrase and grammar-check author-written text. After using this tool, the author(s) reviewed and edited the content as needed and take(s) full responsibility for the content of the published article.

\section*{Conflict of Interest Statement}

The authors declare that they have no known competing financial interests or personal relationships that could have appeared to influence the work reported in this paper.

\section*{CRediT Authorship Contribution Statement}

\textbf{Marium M. Mou:}
Conceptualization, Methodology, Software, Validation, Formal analysis,
Investigation, Data curation, Visualization, Writing -- original draft,
Writing -- review \& editing.
\textbf{Caleb Schenck:}
Investigation, Data curation, Microscopy.
\textbf{Samuel E. Daigle:}
Investigation, Data curation, Formal analysis,
Writing -- review \& editing.
\textbf{William G. Fahrenholtz:}
 Resources, Experimental sample provision, Supervision, Writing -- review \& editing.
\textbf{Bharat Gwalani:}
Investigation, Microscopy, Supervision, Writing -- review \& editing.
\textbf{Stefano Curtarolo:}
Project administration,
Funding acquisition, Supervision, Writing -- review \& editing.
\textbf{Donald W. Brenner:}
Conceptualization, Methodology, Supervision, Project administration,
Funding acquisition, Writing -- original draft, Writing -- review \& editing.

\section*{Data Availability}

The Python scripts used for MC--MD workflow are openly available at this \href{https://github.com/mariummou/HEC_GB_MCMD_MACE_scripts}{repository}~\cite{Mou2026FiniteTemperature}.   The MACE potential used in this work was the publicly available pretrained MACE--OMAT--0 model and can be found in their foundation model repository.

\end{document}